# Helical Luttinger liquid on the edge of a 2-dimensional topological antiferromagnet

Yang Feng[1#†], Jinjiang Zhu[1†], Weiyan Lin[1,2†], Zichen Lian[3†], Yongchao Wang[4], Hao Li[5,6], Hongxu Yao[1], Qiushi He[1], Yinping Pan[1], Yang Wu[6,7], Jinsong Zhang[3,8], Yayu Wang[3,8], Xiaodong Zhou[2,9,10*], Jian Shen[1,2,9,10,11,12*], and Yihua Wang[1,12*]

[1]State Key Laboratory of Surface Physics and Department of Physics, Fudan University, Shanghai 200433, China
[2]Institute for Nanoelectronic Devices and Quantum Computing, Fudan University, Shanghai 200433, China.
[3]State Key Laboratory of Low Dimensional Quantum Physics, Department of Physics, Tsinghua University, Beijing 100084, P. R. China.
[4]Beijing Innovation Center for Future Chips, Tsinghua University, Beijing 100084, P. R. China.
[5]School of Materials Science and Engineering, Tsinghua University, Beijing, 100084, P. R. China.
[6]Tsinghua-Foxconn Nanotechnology Research Center, Department of Physics, Tsinghua University, Beijing 100084, P. R. China.
[7]Department of Mechanical Engineering, Tsinghua University, Beijing 100084, P. R. China.
[8]Frontier Science Center for Quantum Information, Beijing 100084, P. R. China.
[9]Zhangjiang Fudan International Innovation Center, Fudan University, Shanghai 200433, China.
[10]Shanghai Qi Zhi Institute, Shanghai 200232, China
[11]Collaborative Innovation Center of Advanced Microstructures, Nanjing, China.
[12]Shanghai Research Center for Quantum Sciences, Shanghai 201315, China.
[#]Present address: Beijing Academy of Quantum Information Sciences, Beijing 100193, P. R. China.
[†]These authors contributed equally to this work.
[*]Email: zhouxd@fudan.edu.cn; shenj5494@fudan.edu.cn; wangyhv@fudan.edu.cn

**Abstract**

**Boundary helical Luttinger liquid (HLL) with broken bulk time-reversal symmetry belongs to a unique topological class which may occur in antiferromagnets (AFM). Here, we search for signatures of HLL on the edge of a recently discovered topological AFM, MnBi$_2$Te$_4$ even-layer. Using scanning superconducting quantum interference device, we directly image helical edge current in the AFM ground state appearing at its charge neutral point. Such helical edge state accompanies an insulating bulk which is topologically distinct from the ferromagnetic Chern insulator phase as revealed in a magnetic field driven quantum phase transition. The edge conductance of the AFM order follows a power-law as a function of temperature and source-drain bias which serves as strong evidence for HLL. Such HLL scaling is robust at finite fields below the quantum critical point. The observed HLL in a layered AFM semiconductor represents a highly tunable topological matter compatible with future spintronics and quantum computation.**

HLL is a unique state of matter in one-dimension (1D) [1]. 1D electron systems cannot be described by Landau's Fermi liquid theory widely used in higher dimensions because even weak interactions induce strong perturbations. Instead, Tomonaga-Luttinger liquid theory is a proper theoretical framework for a 1D quantum liquid [2,3]. The spin degree of freedom plays paramount importance to its emergent behavior as quantum confinement strictly limits the channels of interaction which depend on the spin. HLL is one type of 1D quantum liquid categorized by the underlying symmetry of excitations (Fig. 1a).

A non-chiral spinless liquid is the case where the electrons are not spin-polarized and conventional Luttinger liquid behavior is expected in charge transport in systems with weak spin-orbit coupling [4-7]. If the spin-orbit coupling is strong, HLL with even number of time-reversal invariant pairs (even HLL) occurs. Due to the fermion doubling theorem [8], however, a pure 1D lattice cannot host an odd HLL (with odd number of time-reversal invariant pairs), which only arises as a metallic edge of a 2D bulk with topological band structure (Fig. 1b). The chiral Luttinger liquid as 'half' of an odd HLL also requires a topologically non-trivial 2D bulk, such as on fractional quantum Hall (QH) edges [9–11]. A Fermi liquid to Luttinger liquid transition was also observed from the QH to the fractional QH regime [12]. The odd HLL is present at the edge of quantum spin Hall insulators [13–19], which is protected by TRS. Breaking TRS by applying a finite field destroys the bulk topological order and consequently the HLL on its edge [20]. In contrast, a TRS-broken bulk hosting odd HLL on its edge belongs to a different topological class which may be robust against a finite field as long as the underlying symmetry is preserved.

In search for HLL with broken TRS, antiferromagnets (AFM) with non-trivial band-structure are promising candidates. This is because a combined space-time (PT) symmetry in AFM, where P denotes the lattice translation and T denotes the time-reversal, can lead to a $Z_2$ topological classification [21] which distinguishes AFM topological insulator (TI) from an ordinary insulator. Such PT symmetry naturally breaks TRS and is robust against a finite field less than the spin-flop field of the AFM order. Due to the vast variety of magnetic structures with vanishing magnetization at the ground state, the symmetry-sensitive topological phases in AFM are extremely rich: AFM TI [21–23], axion insulator [24–26], Dirac semimetal [27–29], AFM Chern insulator [30] and Dirac nodal line metal [31]. Despite the increasing recognition of potential interplay between magnetism, topology and emergent phases at reduced dimension, HLL with broken TRS remained undetected due to the lack of layered topological AFM materials and sensitive techniques to directly identify it.

The recent discovery of the stoichiometric compound $MnBi_2Te_4$ [32,33] offers us an AFM semiconductor with a topological band structure. It is composed of van der Waals stacking of Te-Bi-Te-Mn-Te-Bi-Te septuple layers (SL), making it suitable for fabricating 2D structures with various thickness. Local moments of $Mn^{2+}$ order ferromagnetically within each SL with an out-of-plane easy axis, and antiferromagnetically between SLs. As a result, even or odd number of SLs in thin crystals has different spin symmetry and distinctive topological orders (Fig. 1c) [25]. The odd layer crystals realize a quantum anomalous Hall ground state with Chern number $C = 1$ [34]. For the even layers, on the other hand, two distinctive topological phases with $C = 0$ are possible. If all the surface states, including those on the sides, are fully gaped by either the magnetization normal to the surface (Fig. 1c) or other extrinsic disorder breaking PT symmetry, an axion insulator is predicted [25,26] [35]. However, if only the top and bottom surfaces are gapped by the out-of-plane magnetization, while the AFM side surfaces preserve the PT symmetry and retain the gapless surface states [26,36–38], a TRS-broken AFM TI phase with odd HLL on its edge may occur. In this work, we show evidence for HLL by imaging the helical edge current in the AFM ground state of 6-SL samples. This is achieved by the combination of scanning superconducting quantum interference device (sSQUID) microscopy [39–41] and magneto-transport.

The exfoliated 6-SL sample is tunable to the charge neutral point by applying a back gate voltage ($V_g$). When all the local moments are aligned by a large external field (Fig. 1d), the Hall resistivity $\rho_{yx}$ exhibited a transition from the $C = 0$ zero Hall plateau (Fig. 1e and f, light green shaded) to the $C = 1$ quantized Hall plateau (light blue shaded). This is a result of the magnetic order switching from a compensated AFM ground state to a fully polarized ferromagnetic state under high field [34,35] and serves as a clear indication of the even layer number assignment of this high quality sample.

We first show that the AFM ground state at zero Hall plateau is a topologically distinct phase from the Chern insulator at high fields. Here we adopt a special setup (Fig. 1g) to probe the bulk resistance $R_{bulk}$ so that the contribution from edge carriers is excluded. The current flows through only a pair of electrodes H and C on opposite sides of the device while all the other electrodes are grounded. In this setup, current flowing along the edge does not enter the ammeter and we can obtain $R_{bulk} = V/I_{bulk}$. $R_{bulk}$ as a function of $\mu H$ at different $T$ (Fig. 1h) reveals two metal-insulator transitions as the curves cross two fixed points. The two metal-insulator transitions are evident from the temperature dependence of $R_{bulk}$ at selected fields as well (Figs. 1i and j). Three field regimes are identified: (1) the $C = 0$ insulator phase at $\mu_0 H < 3.98$ T, (2) the metallic spin-flop phase where $R_{bulk} \propto T$ in the intermediate 3.98 T $< \mu_0 H < 4.75$ T, and (3) the Chern insulator phase at $\mu_0 H > 4.75$ T. The metallic state separating the two insulating phases strongly suggests that the bulk bandgap closes and re-opens as the field is reduced. Therefore, the low field order with AFM structure is topologically different from the ferromagnetic Chern insulator.

Having shown the unique topological order in the AFM ground state of the bulk, we now turn to its edge channels by employing the sSQUID to image the current density distribution [16,17]. We apply current between electrodes E and H (Fig. 2a) and scan in the square area (Fig. 2a, light blue dotted). The current flux images (Figs. 2b and c) show a change from bulk to edge conduction at $V_g = 30\ V$. When the charge transport is bulk dominated at $V_g = 0$ V (Fig. 2b), the current flux profile crosses smoothly through zero inside the Hall bar, indicating a homogeneous current flow through the device. As the Fermi level is shifted into the bulk gap at $V_g = 30$ V, we observe a clear shift of current flux distribution (Fig. 2c) towards the top edge of the Hall bar as well as its left edge. It bypasses the broken area between electrode A and B (Fig. S1). The reconstructed current density distributions (Fig. 2d and e) obtained from the current flux images (Figs. 2b and c) [42] are consistent with the above analysis. The absence of the edge current on the broken side suggests the edge transport is suppressed by strong disorder, which were likely introduced from an electrostatic discharge. The current density distribution when current is biased through a different pair of electrodes shows similar edge transport at $V_g = 30$ V (Fig. S4), further confirming the existence of edge current. Therefore, these sSQUID current images provide a direct evidence of edge states inside the bulk gap.

We then perform continuous gate-voltage tuning to follow the evolution of in-gap edge states. The current is applied from top to bottom in this measurement so that it can be directly compared with transport (Fig. 2f, electrodes F to K). The single-line scan of current flux (along the blue dotted line in Fig. 2f) is displayed in Fig. 2g as a function of $V_g$. When $V_g = 0$ V, the zero-flux point appears at the center of the device, indicating a uniform current distribution through the bulk of the device. As $V_g$ increases, the zero-flux point moves to the left edge of the device, indicating a shift of current density towards the edge. The overall trend of current flux distribution as a function of $V_g$ is qualitatively consistent with the longitudinal resistance ($R_{xx}$) measured under the same configuration (Fig. 2h). Under zero-field, $R_{xx}$ increases as $V_g$ tunes the chemical potential from the bulk into the bulk gap; whereas the trend is the opposite under $\pm 9$ T (Fig. 2h). In the charge neutral region (15 V $< V_g <$ 35

V), the large $R_{xx}$ at 0 T in contrast to the diminishing $R_{xx}$ at $\pm 9$ T suggests the edge channel in the AFM state is characteristically different from the chiral edge state of the Chern insulator [36,37].

There are arguments for the existence of impurity phases in MnBi$_2$Te$_4$ acting like Mn-doped Bi$_2$Te$_3$ which may generate chiral edge states for even layers at zero-field [43]. We can differentiate chiral from helical edge states in sSQUID by reversing the direction of the applied current and image the symmetry of its path. Chiral edge current flows around the sample with the same circulating direction after reversing the current bias (Fig. 3a), whereas helical edge current observes TRS and should show antisymmetric current distribution (Fig. 3b). Considering the edge between electrodes A and B are broken in our device (Fig. S1) due to electro-static discharge, if current is only allowed to flow along the lower edge through A and B, the chiral edge channel will be totally blocked, forcing the current to flow inside the bulk (Fig. 3c). However, in the helical case (Fig. 3d), the current flow will not be blocked under opposite bias current since a pair of counterpropagating edge states carries the charge and there always exists an edge mode in the upper edge of the device regardless of the bias. In our device, we do not observe any imbalance of current flux density in the bulk with reversed bias current flow in the device (Figs. 3e and f). Such symmetric current flow under time-reversal is reproducible on different even-layer devices independent of edge contacts (Fig. S11). Comparing the current flux along dotted lines in Figures 3e and 3f, we find the blue-red and green-orange linecut pairs are symmetric (Fig. 3g, grey line $\Phi = 0$). The reconstructed current distribution images (Figs. 3h and i) are also identical regardless of the direction of the current. This fact further corroborates the helical nature of the edge states and rules out its origin from the impurity phases.

Having shown the existence of helical edge states in the AFM ground state of 6-SL MnBi$_2$Te$_4$, we investigate the dimensionality and emergent behavior of the edge excitations. A wide stripe of enhanced local microwave conductivity on the edge of an even-layer sample [37] is technique-specific and is not a true reflection of the intrinsic width of the edge channel [44]. We resort to a particular transport configuration (Fig. 4a) that distinguishes the edge excitation from the bulk [36,45] to overcome the resolution limit in imaging. Under this configuration, only a pair of the nearest electrodes are used and all the others grounded. Recalling that the $R_{\text{bulk}}$ inside the bandgap at zero-field is larger than 1 MΩ (Fig. 1h), the dominant contribution in this two-terminal measurement comes from the edge channels. When the excitation currents are small ($< 5$ nA), we obtain a good straight line on the log-log plot of the conductance versus temperature ($G(T)$), indicating a power-law dependence $G \propto T^\alpha$ (Fig. 4b). We take $I = 3$ nA as an example to get the exponent $\alpha \approx 0.313$. As the excitation current increases, it deviates from such a power-law below a critical temperature $T_c$. The overall behavior of $G(T)$ is reminiscent of the Luttinger liquid behavior for a 1D electron system with internal tunneling process [2,3]. In stark contrast, the scaling exponent in the Chern insulator phase is close to 0 (Fig. 4c), which is consistent with excitations of a Fermi liquid as explained below.

For a helical edge state, two-particle backscattering is allowed according to Kramers theorem. Two right movers excited near the Fermi surface can be mixed with two left movers. The 'inter-branch' scattering creates electron-hole pairs, which are bosonic. Such 'inter-branch' scattering is responsible for the HLL behavior leading to the famous spin-charge separation [1,15]. On the other hand, a chiral edge lacks the Kramers pair for 'inter-branch' scattering. Instead, the 'intra-branch' scattering within a chiral edge creates an electron-hole pair and simultaneously annihilates another near the Fermi surface. This process only renormalizes the velocity of charge and does not change its direction. Therefore, for a Chern insulator with integer quantum Hall conductance, the non-interacting 1D edge acts as a Fermi liquid, in which $G = dI/dV$ remains constant at elevated bias $V$ [12].

To obtain the Luttinger parameter ($K$) of the HLL, we measure the bias voltage $V_{DC}$ dependence of $G$ at different $T$ (Fig. 4d). $G$ is independent of the bias when $eV \ll k_B T$, but follows a power law with the fitted exponent of 0.167 when $eV \gg k_B T$. The crossover from linear (Ohmic) to power law scaling with $T$ is a hallmark of HLL theory [46,47,4], which can be captured by the expression:

$$G \propto T^\alpha \cosh\left(\gamma \frac{eV}{2k_B T}\right) \left| \Gamma\left(\frac{1+\alpha}{2} + \gamma \frac{ieV}{2\pi k_B T}\right) \right|^2$$

where $\gamma$ is a constant characterizing the largest internal tunneling barrier, and $\Gamma(x)$ is the gamma function. We examine the scaling relation $G/T^\alpha \propto eV/k_B T$ by plotting the scaled conductance $G/T^\alpha$ against $eV/k_B T$ (Fig. 4e). The data collapse quite well onto a universal curve over the entire bias and $T$ range with $\alpha = 0.313$. The exponent of these powers depends on the number of 1D channels and is related to $K$ characterizing the sign and the strength of the electron-electron (e-e) interactions. According to the perturbation theory, under strong Coulomb interaction corresponding to $K < K_c = 1/4$, the HLL is gapped at zero $T$ for a quench disorder-induced random but relevant two-particle backscattering [1,48]. At low but finite $T$, $G$ is restored by half-charge tunneling [49] between two energy minima inside the HLL, with power-law of $G(T) \propto T^{2(1/4K-1)}$. Following the relation $\alpha = 2(1/4K - 1)$, we obtain $K = 0.216$, indicating repulsive interactions [49,50]. As a comparison, $K$ is 0.2 in InAs/GaSb heterostructure [18] but reached 0.5~0.8 in HgTe quantum wells from simulations [49,51]. These scaling analysis provides a critical evidence of the internal tunneling processes [1,49] between branches of HLL in 6-SL device (Fig. 4b).

The side surface of a 6-SL flake will open a significant bandgap ~ 0.047 eV due to quantum confinement in its thickness [26]. This prevents the side surface from forming a 1D conducting channel. Surface Hall effect [52,53] in the axion phase of $MnBi_2Te_4$ [54,55] cannot explain our observation either: earlier simulations showed that the minimum number of SLs needed for half quantized chiral currents is about 100 [ [54]], which is much larger than that of our device. In short, 1D helical edges induced by the hybridization of the top and bottom surfaces in AFM [56] is the only reasonable scenario consistent with all of our observations.

To understand the origin of the HLL, we consider four low-lying states without spin-orbit coupling or exchange effect, denoted by $|+\uparrow\rangle$, $|-\uparrow\rangle$, $|+\downarrow\rangle$ and $|-\downarrow\rangle$ corresponding to opposite parity and spin respectively [25,57]. This base leads to an effective Dirac Hamiltonian of AFM MnBi$_2$Te$_4$:

$$H = \begin{pmatrix} m & A_2 k_z & 0 & A_1 k_- \\ A_2 k_z & -m & A_1 k_- & 0 \\ 0 & A_1 k_+ & +m & -A_2 k_z \\ A_1 k_+ & 0 & -A_2 k_z & -m \end{pmatrix} + M\sigma_z \otimes I$$

where $k_\pm = k_x \pm i k_y$, the mass term $m = m_0 - B_1(k_x^2 + k_y^2) - B_2 k_z^2$, $I$ is the identity matrix in the parity subspace and $M$ is the average magnetization. Due to the quantum confinement, the quantum well states lead to 2D sub-band series which are still inverted for 6-SL [25]. This effective model captures topologically distinct bulk band structures of 6-SL MnBi$_2$Te$_4$ under different magnetic orders (Fig. 4f). At the low field AFM regime, $M = 0$ and this model is similar to the Bernevig-Hugh-Zhang model for the HgTe quantum wells [13]. The even parity bands $|+\rangle$ (Fig. 4f, AFM, solid) and odd parity bands $|-\rangle$ (dashed) are both inverted. They hybridize and generate a gap (pink and purple), leading to a pair of helical edge states (Fig. 4b, inset). In the quantum critical regime (Fig. 4f, SF), the increasing exchange field leads to finite $M$, which strengthens the band inversion between $|+\downarrow\rangle$ and $|-\uparrow\rangle$ but releases the band inversion between $|+\uparrow\rangle$ and $|-\downarrow\rangle$ (Fig. 4f, SF, pink) so that their bandgap closes and the ground state becomes metallic. In the high field ferromagnetic regime (Fig. 4f, FM), $M$ is large enough to reopen the bandgap between $|+\uparrow\rangle$ and $|-\downarrow\rangle$ as the system becomes a Chern insulator. We note that the suppression of the helical edge state by strong disorder, such as the case we observed on the lower-right edge of the sample, is not contradictory to the effective model [25, 52].

At last, we examine the robustness of the edge HLL under finite magnetic field. We extract the scaling exponent as a function of field for another 6-SL sample (Fig. 4f). $\alpha$ stayed around 0.4 in the AFM regime within $\pm 3$ T. Upon entering the quantum critical regime (as can be seen from $R_{yx}$), $\alpha$ first rises and then quickly decreases at $\pm 4$ T. This reduction in $\alpha$ is concurrent with the rise of $R_{yx}$ to $\sim 0.9\ h/e^2$. Increasing the field to the ferromagnetic regime, the system enters the Chern insulator, for which tunneling experiments is more appropriate for obtaining a reliable $\alpha$. Nevertheless, the stable $\alpha$ in the AFM regime strongly suggests that the HLL we observed in the TRS-broken bulk is robust against a finite magnetic field not exceeding its AFM spin-flop field. In contrast, the odd HLL in a TRS quantum spin Hall insulator is unstable under a magnetic field [20], which is a direct consequence of breaking the TRS protection of its band topology. The robust odd HLL on the side surface of an AFM TI provides a promising 1D system to host more exotic quantum matter [21,38,58,59,60,61]

In conclusion, we combine sSQUID and transport measurement to identify an odd HLL on the edge of 6-SL MnBi$_2$Te$_4$. Its insulating bulk at low field is a topological AFM which breaks TRS but is distinct from the Chern insulator in the ferromagnetic order. We further show that the HLL is robust against finite magnetic field as it originates from its TRS-broken bulk bands. Our observations demonstrate the edge of even layer MnBi$_2$Te$_4$ as a tunable 1D platform which may find applications in spintronics and topological quantum information.

## Acknowledgement

YHW would like to acknowledge partial support by National Natural Science Foundation of China (Grant No. 11827805 and 12150003), National Key R&D Program of China (Grant No. 2021YFA1400100) and Shanghai Municipal Science and Technology Major Project (Grant No. 2019SHZDZX01). The work at Tsinghua University is supported by National Natural Science Foundation of China (Grant No. 21975140, and No. 51991313), the Basic Science Center Project of National Natural Science Foundation of China (Grand No. 51788104) and the National Key R&D Program of China (Grand No. 2018YFA0307100). Y. Feng would like to acknowledge support by the NSFC Grant No. 11904053, National Postdoctoral Program for Innovative Talents (Grant No. BX20180079) and China Postdoctoral Science Foundation (Grant No. 2018M641904). X. D. Zhou would like to acknowledge partial support by the NSFC Grant No. 12074080, and Shanghai Science and Technology Committee Rising-Star Program (Grant No. 19QA1401000). All the authors are grateful for the experimental assistance by T. Zhang and S. Y. Li, and for the stimulating discussions with W. Ruan, C. Z. Chen, C. J. Wu and J. Wang.

# Figures

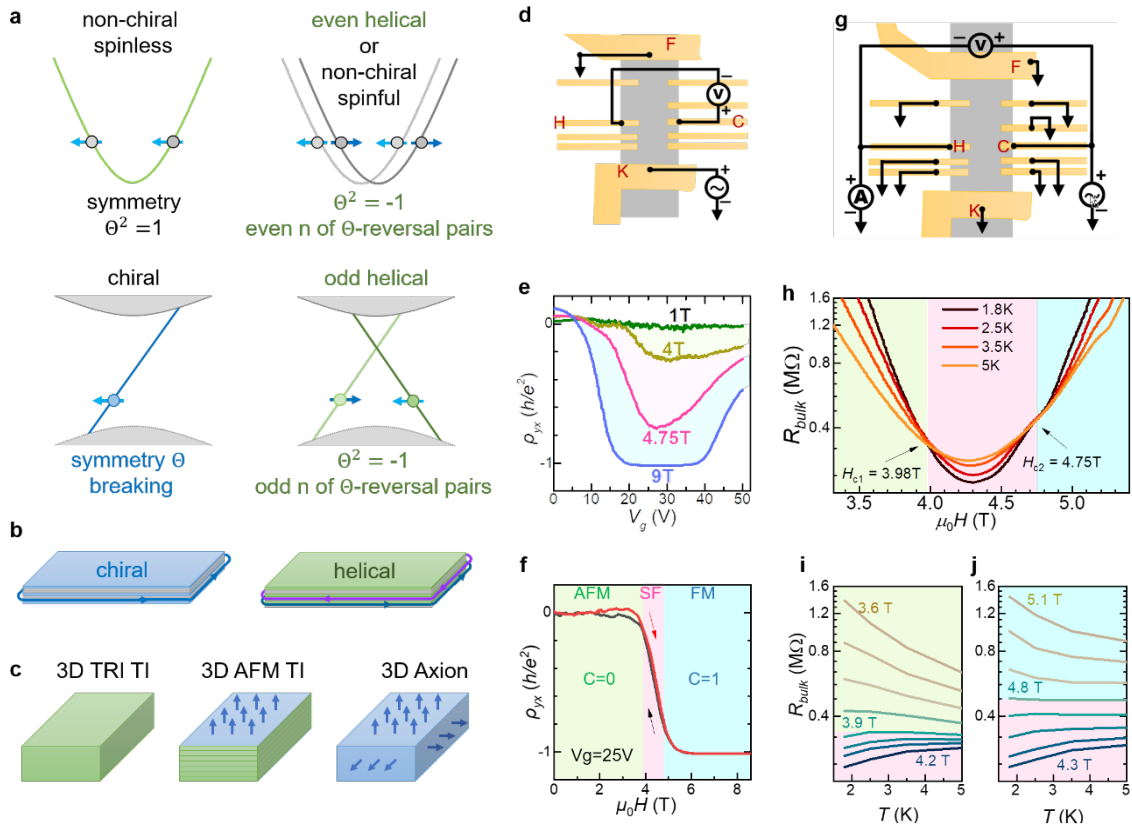

**Fig. 1 | Magnetic field driven quantum phase transition (QPT) in a 6-septuple-layers (6-SL) MnBi$_2$Te$_4$ flake. a,** Illustration of band dispersions for the family of Luttinger liquids. Non-chiral spinless Luttinger liquid satisfies time-reversal (TR) transformation $\Theta^2 = 1$. Even helical Luttinger liquid (HLL) has even number ($n$) of TR pairs with $\Theta^2 = -1$. Chiral Luttinger liquid breaks time-reversal symmetry (TRS). Odd HLL has odd $n$ of TR pairs with $\Theta^2 = -1$. **b**, Schematic drawing of bulk 2D lattices hosting chiral Luttinger liquid and HLL, respectively. **c**, Family of symmetry-protected topological states in 3D lattices. Time-reversal invariant (TRI) topological insulator (TI) has gapless surfaces protected by TRS. Antiferromagnetic (AFM) TI breaks TRS but has gapless side surfaces protected by space-time symmetry. 3D axion insulator does not have any gapless surfaces. Through **a-c**, the green areas denote symmetry-protected surface states, while the blue areas denote surface states gaped by magnetic ordering. **d,** Measurement diagram for Hall resistivity $\rho_{yx}$. **e,** gate voltage $V_g$ dependence of $\rho_{yx}$ at different fields $\mu_0 H$. **f,** $\mu_0 H$ dependence of $\rho_{yx}$ showing QPT. The magnetic states are indicated on the plot (SF, spin-flop; FM, ferromagnetic). **g,** The setup measuring bulk resistance $R_{\text{bulk}}$ using electrodes C and H with all the other electrodes grounded. **h,** $\mu_0 H$ dependence of $R_{\text{bulk}}$ in the QPT regime at different temperatures $T$. Black arrows indicate critical points at $H_{c1} = 3.98$ T and $H_{c2} = 4.75$ T. **i** and **j,** $T$ dependence of $R_{bulk}$ around $H_{c1}$ and $H_{c2}$, respectively. Throughout **e-j**, the pink, light green and light blue shaded areas denote the bulk-insulating AFM with Chern number C = 0, metallic SF and the bulk-insulating FM with C = 1, respectively.

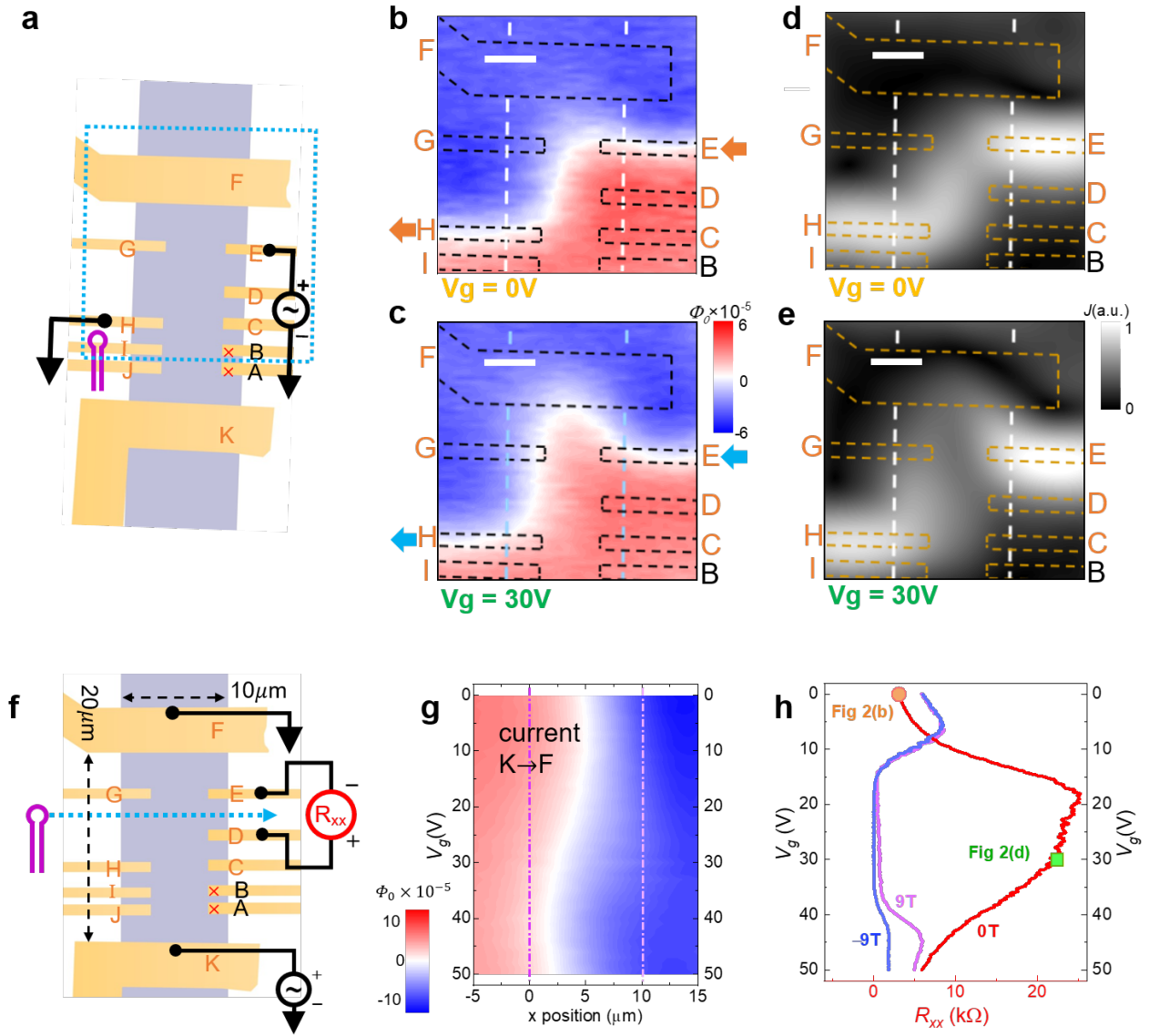

**Fig. 2 | Scanning SQUID imaging of edge current tuned by gate voltage $V_g$. a,** Schematic diagram for current imaging. Current flows from electrode E to H. Light blue dotted line indicates the square area where scans are taken. The purple solid ring represents the relative orientation and size of the pick-up coil of nano-SQUID to the device. The blue arrow with dotted line indicates the direction of single line scans. **b** and **c,** current flux images at $V_g$ = 0 V and 30 V, respectively. The dashed lines outline the approximate geometry of the sample. **d** and **e,** Reconstructed 2D current density from **b** and **c,** respectively. Scale bars are 5 μm. **f,** measurement diagram of the Hall bar device for line scans. Electrodes A and B are broken before the scanning measurement. **g,** The line scans along the blue dotted line in **a** as a function of $V_g$. The dash-dotted lines indicate the positions of device edges. **h,** $V_g$ dependence of $R_{xx}$ measured between D and E as shown in **f**. The red, blue and purple color denote $\mu_0 H$ = 0 T, -9 T, +9 T, respectively. The orange circle and green square indicate the values of $V_g$ where current flux images are taken.

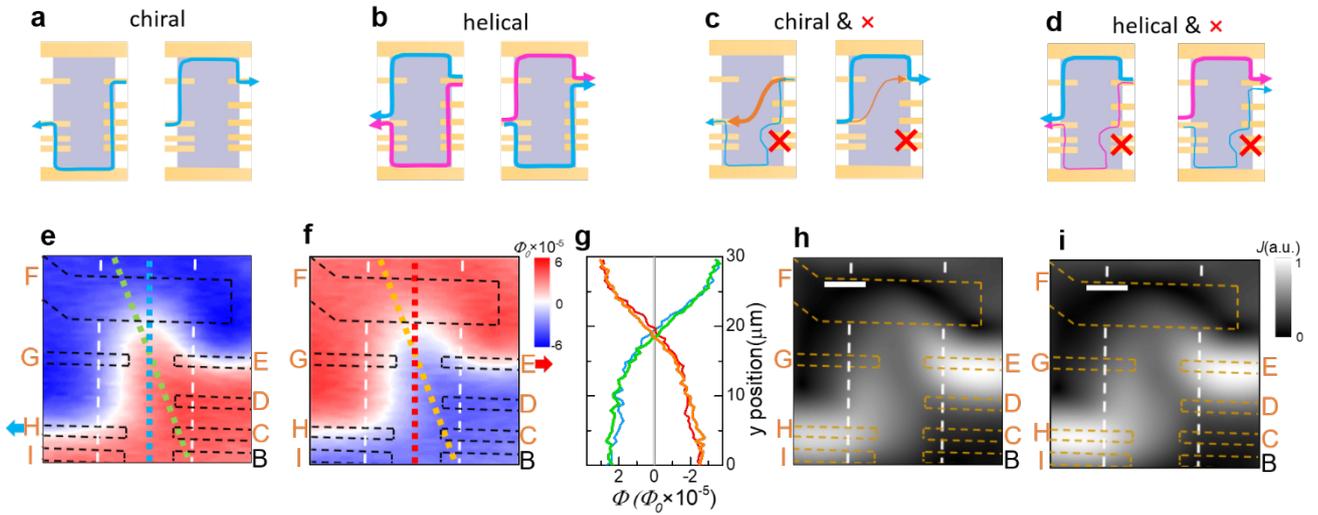

**Fig. 3 | Helical edge current under time reversal measurement. a** and **b,** Illustrations of current distributions of chiral and helical edge states under opposite current bias. The current density is only antisymmetric for the helical case. **c** and **d,** Current distributions when the bottom right electrodes are broken such that the conduction is forbidden along the lower-right edge. In the chiral case, the current will flow through the shortest path inside the bulk for the right-to-left bias, and through the left and top edge for the opposite bias. In the helical case, current density is still antisymmetric even if the edge is partially blocked. **e** and **f,** Current flux images of the multi-terminal Hall bar device under opposite current bias. The flux images are obtained at $V_g = 30$ V for the DC bias current flowing from electrode E to H in **e**, and reversed in **f**. **g,** Analysis of flux under time reversal operation. The flux is extracted along the linecut indicated by the dotted lines in **e** and **f**. **h** and **i,** reconstructed 2D current density. Scale bars are 5 μm.

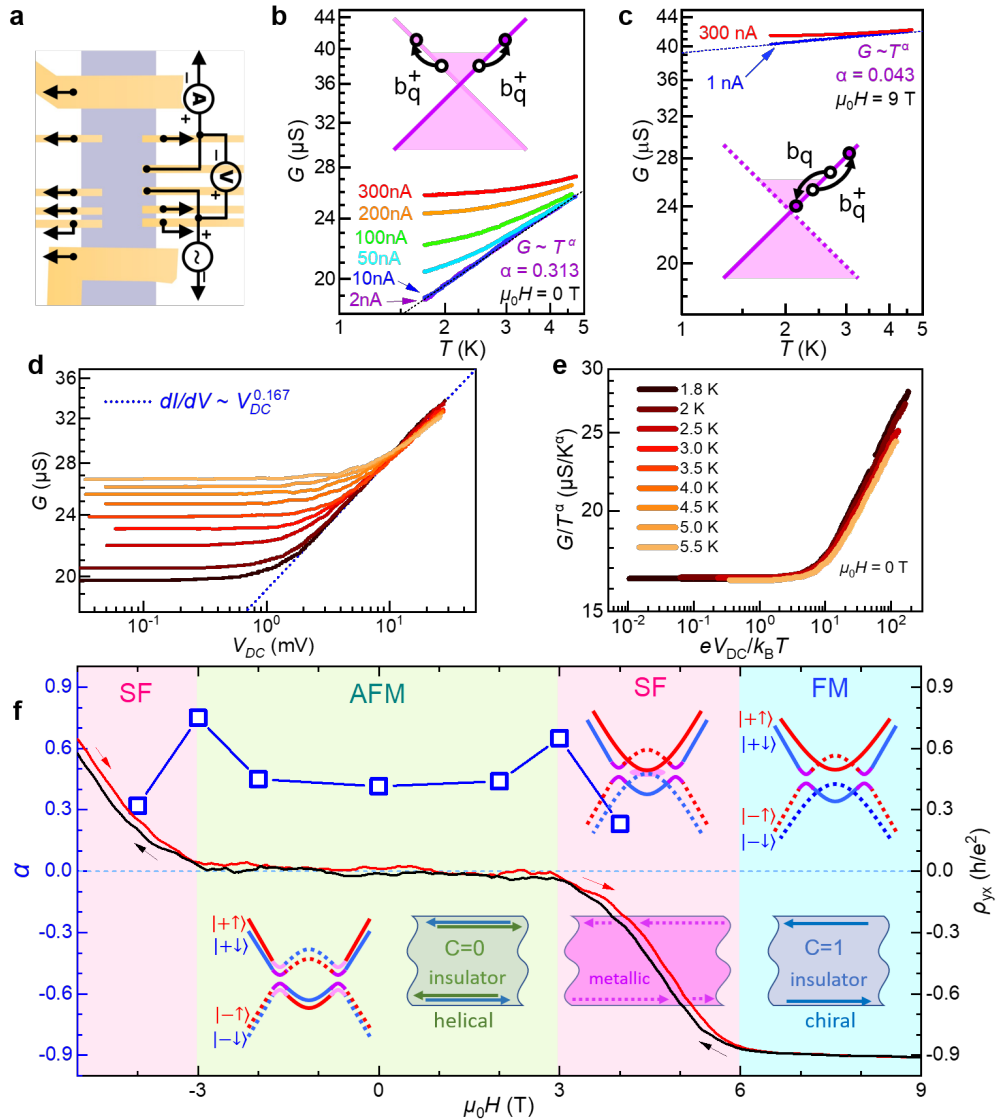

**Fig. 4 | HLL scaling behavior of edge excitations in the AFM order. a,** The setup for the measurement of edge conduction between the nearest pair of electrodes (edge length ~ 2 μm) with all the other electrodes grounded. **b,** Excitation dependence of the helical edge conductance $G$ with $I_{AC}$ = 2 nA to 300 nA at different $T$. The straight dotted line on the log-log plot indicates a power-law behavior $G \propto T^\alpha$ with $\alpha \sim 0.313$. The 'inter-branch' scattering creates electron-hole (e-h) pairs, which can be regarded as bosons. **c,** Similar as **b** but taken at the chiral regime of $\mu_0 H$ = 9 T. The 'intra-branch' scattering within a chiral edge creates an e-h pair which then annihilates near the Fermi surface. Hence it is still a fermi liquid. **d,** $V_{DC}$ dependence of the edge differential conductance $dI/dV$ measured at $T$ = 1.8 K to 5.5 K, with the ac modulation current $I$ = 2 nA. The solid line indicates a power law of $dI/dV \propto V_{DC}^{0.167}$. **e,** The data in **d** replotted using temperature scaling, which shows that all the measured data points except the saturation region collapse onto a single curve. **f,** Left axis shows the extracted $\alpha$ and right axis shows $\rho_{yx}$ as a function of magnetic field. The data was obtained from a second 6-SL MBT sample (SOM). The illustrations in the inset shows a four-band effective model (see text) depicting the evolution of the bulk bands under three different magnetic orderings. The solid and dashed lines denote the subbands with even and odd parity, respectively. The blue and red color denote down and up spins respectively.

In AFM, the initial bands have inverted parity and the crossover-induced gap opening and opposite parity give rise to the helical edge states with $C = 0$. In SF, the exchange field releases the band inversion in one pair of the subbands (red solid and blue dotted), and increases the band inversion in the other pair. This mechanism closes the bulk gap and the system becomes metallic. In FM, the bulk gap reopens and the system enters the Chern insulator state with $C = 1$. The characteristic HLL scaling behavior is robust in the AFM regime.